\def\year{2021}\relax
\documentclass[letterpaper]{article} 
\usepackage{aaai21}  
\usepackage{times}  
\usepackage{helvet} 
\usepackage{courier}  
\usepackage[hyphens]{url}  
\usepackage{graphicx} 
\usepackage{natbib}  
\usepackage{caption} 
\title{Reliability of Content and Echo Chambers on YouTube during the COVID-19 Debate}
\author{
    Niccolò Di Marco\textsuperscript{\rm 1},
    Matteo Cinelli\textsuperscript{\rm 2},
    Walter Quattrociocchi\textsuperscript{\rm 2}
    \\
}
\affiliations{
    \textsuperscript{\rm 1} University of Florence
    \textsuperscript{\rm 2} Sapienza University of Rome\\
}

\begin{document}

\maketitle

\begin{abstract}
The spread of inaccurate and misleading information may alter behaviours and complicate crisis management, especially during an emergency like the COVID-19 pandemic.
This paper aims to investigate information diffusion during the COVID-19 pandemic by evaluating news consumption on YouTube.
First, we analyse more than 2 million users’ engagement with 13,000 videos released by 68 YouTube channels, labelled with a political bias and fact-checking index. Then, we study the relationship between each user’s political preference and their consumption of questionable (i.e., poorly fact-checked) and reliable information.
Our results, quantified using measures from information theory, provide evidence for the existence of echo chambers across two dimensions represented by political bias and the trustworthiness of information channels. We observe that the echo chamber structure cannot be reproduced after properly randomising the users’ interaction patterns. Moreover, we observe a relation between the political bias of users and their tendency to consume highly questionable news.
\end{abstract}

\section{Introduction}
The case of the COVID-19 pandemic made explicit the critical role of information diffusion during critical events~\cite{briand2021infodemics}. 
A relevant example was the massive amount of uncertain information shared by the media to justify the withdrawal of one AstraZeneca vaccine batch, which led to a dramatic lack of trust in it.
In fact, the information ecosystem radically changed with the advent of social media platforms as they implement algorithms and interaction schemes to maximise user engagement. Those algorithms account for users’ preferences and may significantly alter social dynamics and information diffusion~\cite{bakshy2015exposure, cinelli2021echo}.
Therefore, it is crucial to understand how people seek or avoid information and how those decisions affect their behaviour~\cite{sharot2020people} when the news cycle — dominated by the disinter-mediated diffusion of content — significantly alters how information is consumed and reported on.
This corresponds to investigating what is called social contagion, i.e., the spread of ideas, attitudes, norms, or behavioural patterns from individual to individual through social influence, imitation, and conformity. Social contagion depends on users’ attitudes, tendencies, and intentionality. 
Our attention span is limited and feed algorithms might further limit our selection process by suggesting content similar to those we are usually exposed to. Plus, users show a tendency to favour information adhering to their beliefs and join groups formed around a shared narrative, that is, echo chambers~\cite{sunstein2004democracy, garimella2018quantifying, cinelli2021echo}. Echo chambers are environments in which users’ opinion, political leaning, or belief about a topic gets reinforced due to repeated interactions with peers or sources having similar tendencies and attitudes. 
This work follows the definition of echo chambers provided in~\cite{cinelli2021echo} to understand the users’ attention patterns on YouTube during the COVID-19 pandemic. According to some studies, YouTube plays a prominent role in the radicalisation of opinions~\cite{ribeiro2020auditing,hosseinmardi2021examining,feezell2021exploring} and in the diffusion of questionable (i.e., poorly fact-checked) content~\cite{ribeiro2020auditing,cinelli2021dynamics} being one of the most visited online domains and information retrieval platforms. In such a context, the COVID-19 debate on YouTube seems to be a suitable case study for investigating the presence of the echo chamber effect and its relationship with the consumption of questionable information.
The dataset exploited in our study contains 10 millions comments to videos published by 68 prominent news channels on YouTube, in the time window ranging from December 2019 to September 2020. We start by introducing some preliminaries used throughout the article.
Then, we analyse users' engagement, in terms of comments posted by the users on videos produced by YouTube channels with a known political bias and fact-checking score. Finally, we investigate the relationship between users' political preferences and their consumption of questionable and reliable information finding that echo chambers exist on YouTube across the political and fact-checking dimensions.

\section{Data and Methods}


We collected videos using the official YouTube Data API in the period ranging from 2020/01/15 to 2020/09/06, searching for videos that matched a list of keywords selected on the basis of Google Trends’ COVID-19 related queries. Those keywords include the terms: coronavirus, nCov, corona virus,  corona-virus, covid or SARS-CoV. An in-depth search was then performed by crawling the network of related videos as provided by the YouTube algorithm. We filtered the related videos that matched our set of keywords in the title or description from the gathered collection and we collected the comments received by those videos.

From all these videos, we kept only a set of 10,968,002 comments posted by 2,092,817 users on 12,933 videos published from 2019/12/2 to 2020/9/5 by 68 YouTube channels directly linked to news outlets classified by Media-Bias-Fact-Check (MBFC)~\cite{mbfc2014}, an independent fact-checking agency. We specify that all these videos come from channels that posses a political bias and a factual reporting index from MBFC.
MBFC provides such indexes for news outlets and we assumed that they are inherited by their official YouTube channels. The first index provides a score for the channel’s bias in the political dimension (i.e., its political leaning). The second represents the overall factual reporting level of the news published by the channel. Since the concept of factual reporting is very broad, we specify that, according to MBFC, a questionable source “exhibits one or more of the following: extreme bias, consistent promotion of propaganda/conspiracies, poor or no sourcing to credible information, a complete lack of transparency and/or is fake news.”

\begin{figure}[h]
  \centering
  \includegraphics[width=\linewidth]{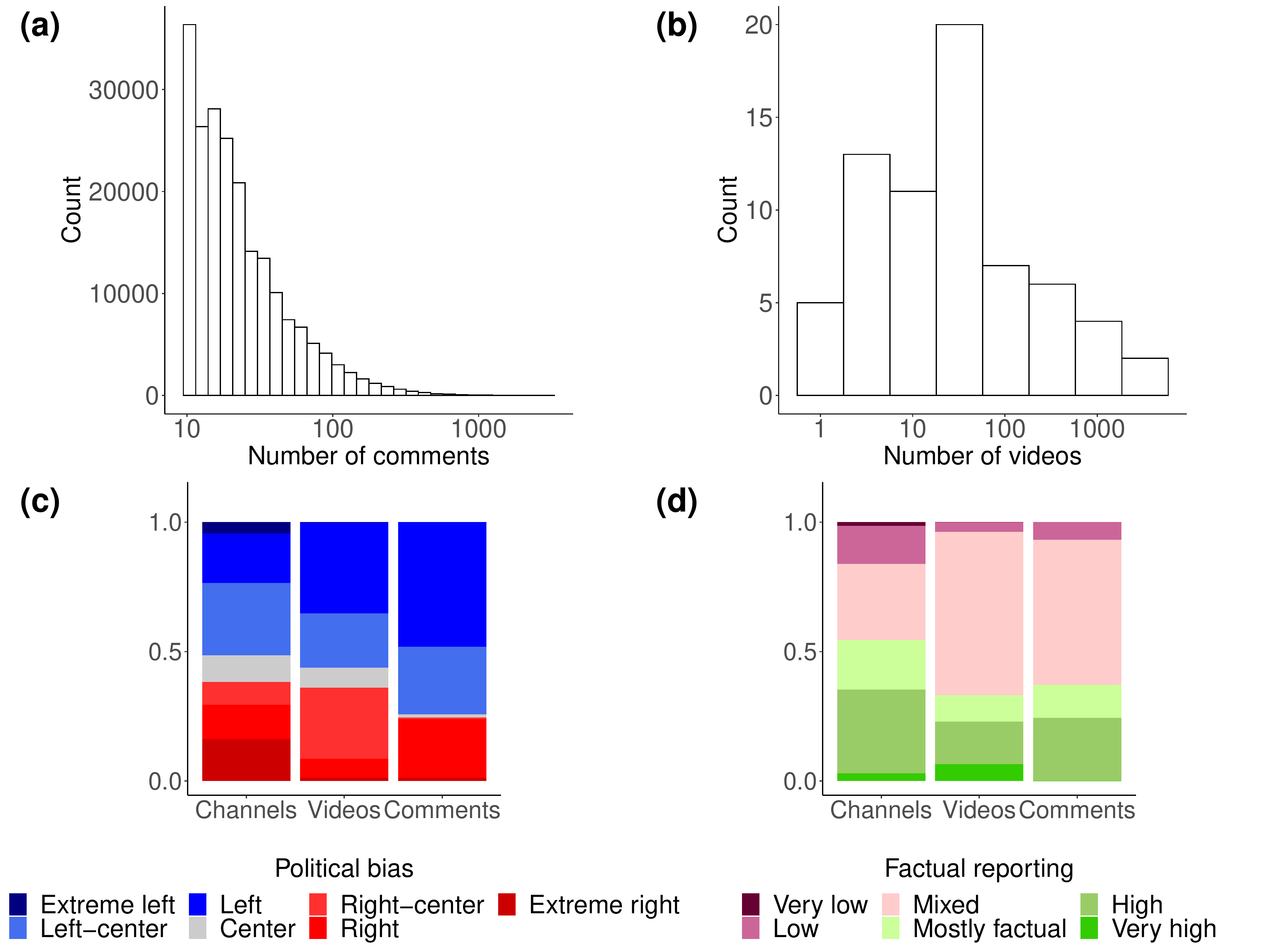}
  \caption{(a) density of comments per user (only users with at least 10 comments were considered), (b) density of videos per channel, (c) proportion of channels, videos and comments considering the political bias of the channels, (d) proportion of channels, videos and comments considering the factual reporting of the channels.}
  \label{fig:dataset_descr}
\end{figure}

For instance, considering the YouTube channel of Breitbart (a popular far-right news outlet), MBFC assigns to it a political leaning corresponding to \textit{Extreme right} and a Factual reporting index corresponding to \textit{Mixed}. Specifically, in the case of Political Bias, we assign numerical values to categorical labels as follows: -1 to Extreme left; -0.66 to Left; -0.33 to Left-center; 0 to Center; 0.33 to Right-center; 0.66 to Right; and 1 to Extreme Right. Accordingly, for the factual reporting, we assign: 0 to Very Low, 0.2 to Low, 0.4 to Mixed, 0.6 to Mostly Factual, 0.8 to High and 1 to Very High. Figure~\ref{fig:dataset_descr} displays some general features of our dataset. In particular, we note that the distribution of videos by channel shows a lognormal-like shape (panel (b)) allowing for large deviations that are actually more likely in the distribution of comments per user (panel (a)), as displayed by the heavy tail of the distribution.
Furthermore, as shown in panels (c) and (d), the labels distributions in the set of channels is (reasonably) uneven and the observed heterogeneity is increasing when we consider first videos and then comments. This means that channels having more frequent labels produce a higher share of videos and receive a disproportionately higher amount of comments. 


We quantify the strength of the preferences of YouTube users through their engagement with videos (specifically the number of comments posted by each user). Consider a user $i$ leaving a total of $n_i$ comments on a set of videos coming from $h_i$ channels, in which channel $j$ has political bias $b_j$. Suppose that $t_j ^i$ is the total number of comments left in channel $j$ by user $i$. We define the political bias of user $i$ as

\begin{equation}\label{pol}
    p_i \equiv \frac{1}{n_i}\sum_{j = 1}^{h_i} t_j^i b_j \, .
\end{equation}

This index represents a (weighted) average of the channels' political leanings on which user $i$ commented and therefore provides information about the user’s political preference/bias. Similarly, each channel has a fact-checking index $f_j$ and we define the persistence index of user $i$ as

\begin{equation}\label{per}
    c_i \equiv \frac{1}{n_i}\sum_{j = 1}^{h_i} t_j^i f_j \,.
\end{equation}

The latter quantity has an interpretation similar to the former. In particular, it is a (weighted) average of the channels' fact-checking indexes on which user $i$ commented and therefore it provides information about the user’s persistence in commenting videos characterised by a certain trustworthiness. 
Finally, note that $p_i \in [-1, 1]$, while $c_i \in [0,1]$.



To study the relationship between the two indexes, $p_i$ and $c_i$, we constructed an unweighted undirected bipartite network $G$ with one partition representing the set of users and on the other the set of channels, and in which two nodes $i$, $j$ are connected if and only if the user $i$ commented at least one video published by the channel $j$. We decided to follow the network approach since it allows us to introduce relational indexes and study the social dynamics typical of a social network, such as YouTube. First of all, we study the interplay between the political bias of users and their persistence indexes by visualising their joint distribution. Then, in order to detect echo chambers, we compute the user bipartite projection $G'$ and we compare the political bias of users with the average political bias of their neighbourhood. $G'$ is a network obtained from $G$: its adjacency matrix is $A = B^t B$, where $B$ is the incidence matrix of $G$. The nodes of $G'$ are the users and two users $i,j$ are connected if and only if they have commented at least one common channel.
We repeat the same analysis for the persistence index.

\section{Results}

\begin{figure}[h]
  \centering
  \includegraphics[width=0.6\linewidth]{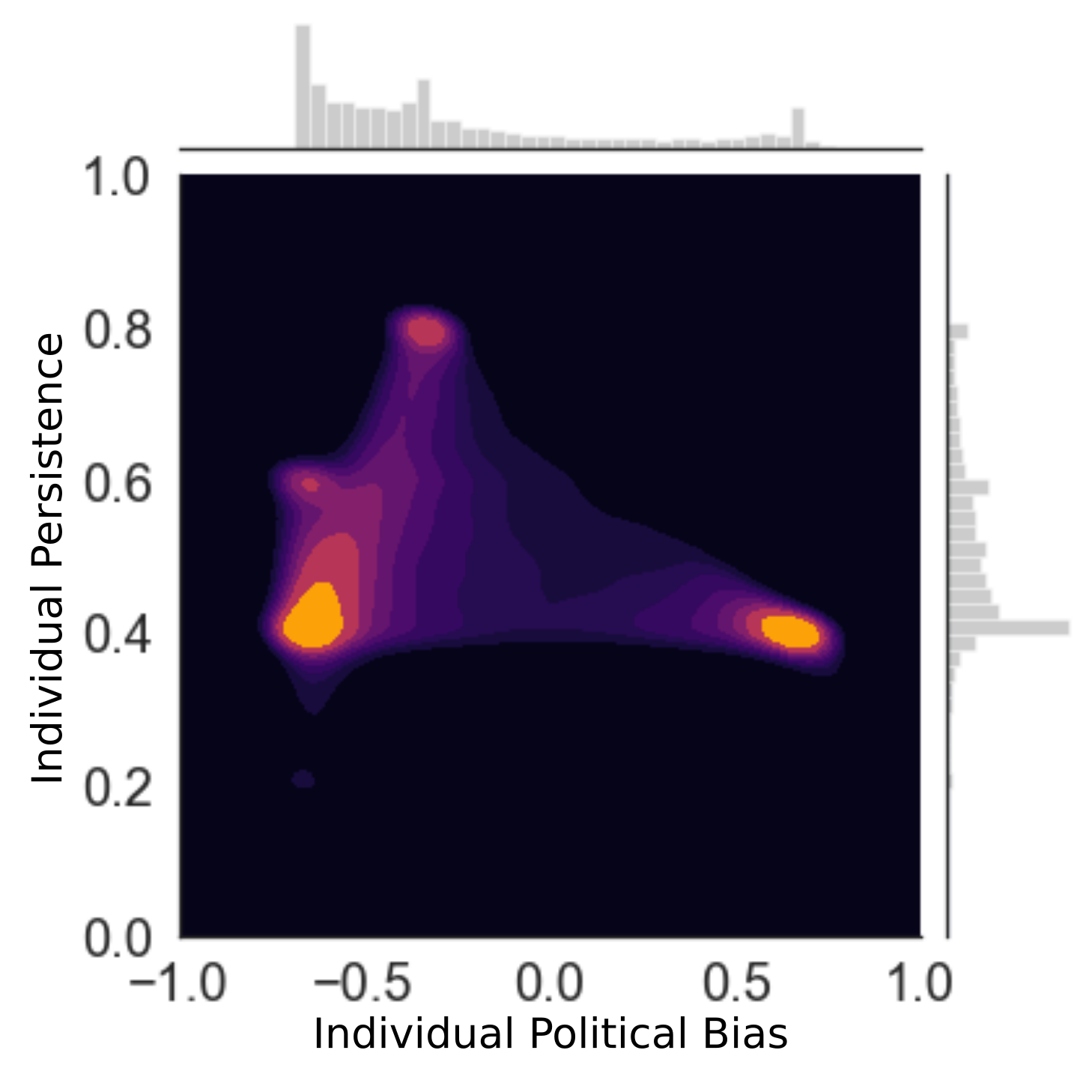}
  \caption{Relationship between the political bias of users with at least 10 comments and their persistence index. Users with a leaning far from the Center tend to consume information from sources with lower fact-checking label. Users with a Left bias display more than one behaviour: a part of them consume more fact-checked news, while the others tend to get information through less reliable channels.}
  \label{fig:bias_pers}
\end{figure}

To test whether a relationship between the users’ political bias and their tendency to consume questionable/reliable content is present on YouTube, we first inspect the joint distribution of the two indexes of bias and persistence, namely $p_i$ and $c_i$.  
In Figure~\ref{fig:bias_pers} we report a 2D density plot showing the relation between the political bias and the persistence of each user. Marginal distributions are also reported. The colour represents the density of users: the lighter, the larger the number of users.
We note a multi-polarisation phenomenon, that is, users tend display an overall opinion focused approximately on three main positions: Left, Left-center and Right. Furthermore, recalling that a higher value of persistence implies a higher factual reporting, we note that users with political leaning far from the Center tend to consume less fact-checked news. Specifically, users that consume content produced by questionable sources are also more likely to have an political leaning skewed towards the extremes. On the other hand, users that consume mostly fact-checked news present, on average, a political bias slightly shifted towards Left. Interestingly, users with a political leaning skewed towards Left display more than one behaviour, namely few of them have a higher persistence score. Simultaneously, a relevant share of users display a low persistence value, possibly indicating that users with a political leaning skewed towards Left comment information from reliable and questionable sources in a somewhat segregated manner.

The echo chamber concept translates into a topological property of the co-commenting network $G'$, in which a user $i$, with a given $p_i$ and $c_i$, is surrounded by other users sharing similar values of such indexes. This concept can be quantified by defining, for each user $i$, the average political leaning and persistence index of its neighbours as follows:

\begin{equation}
    p_i^N = \frac{1}{k_i} \sum_j A_{ij} p_j
\end{equation}

\begin{equation}
    c_i^N = \frac{1}{k_i} \sum_j A_{ij} c_j
\end{equation}

where $k_i$ is the degree of node $i$ and $A_{ij}$ is the adjacency matrix of the user bipartite projection $G'$ obtained from $G$. Specifically, $A_{ij} = 1$ if and only if user $i$ and user $j$ commented at least one common video.

\begin{figure}[h]
  \centering
  \includegraphics[width=\linewidth]{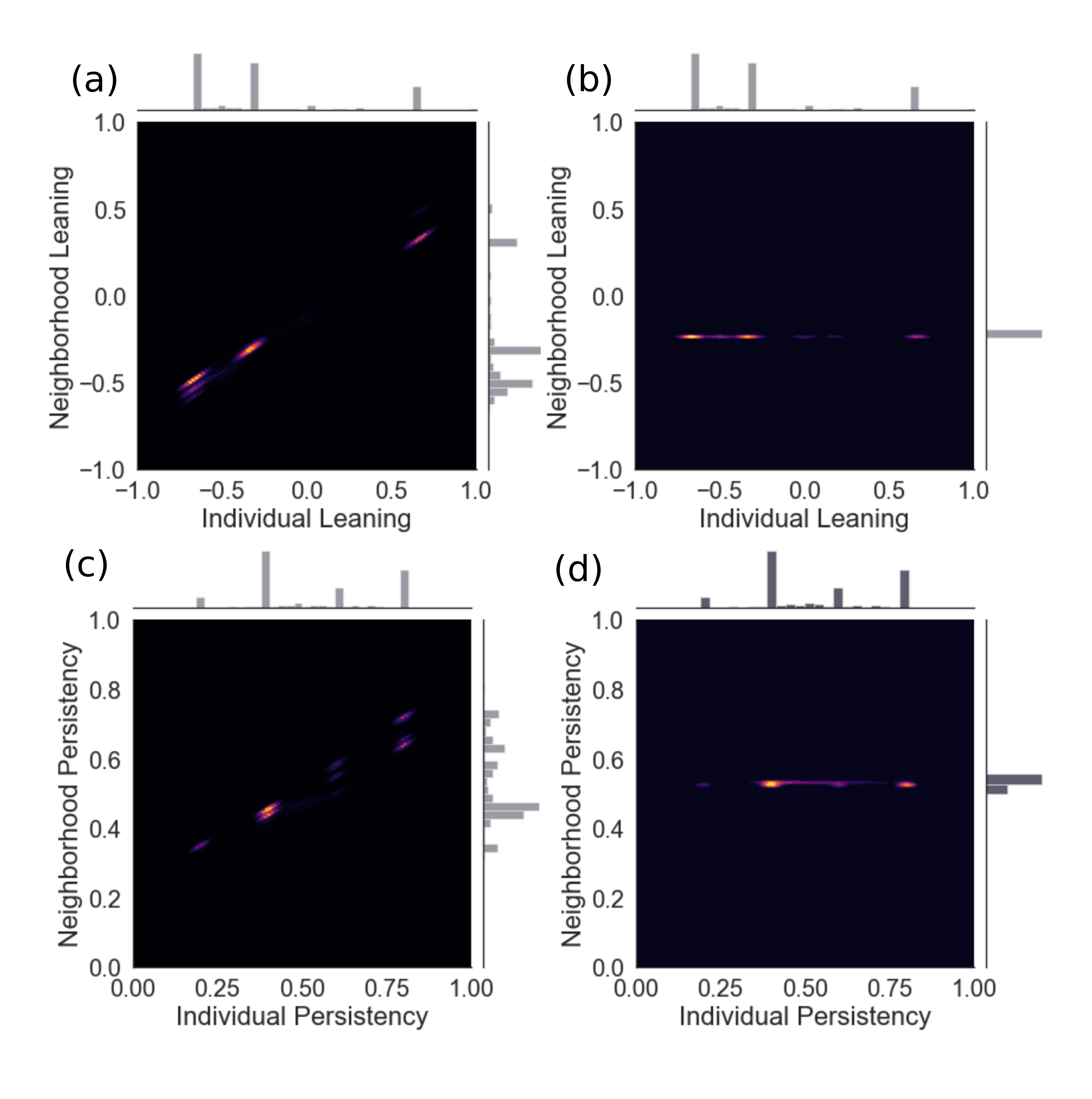}
  \caption{Panel (a) displays the relationship between the political bias of users and the average political bias of their neighbourhood. Panel (c) displays the relationship between the persistence index of users and the average persistence of their neighbourhood. Panels (b) and (d) are obtained analogously to (a) and (c) but in a randomised network.  Panels (a) and (c) show the presence of echo chambers from both the political and the questionable/reliable dimension while panels (b) and (d) confirm that echo chambers do not arise from random behaviour.}
  \label{fig:echo_chambers}
\end{figure}

Figure~\ref{fig:echo_chambers} shows the joint distribution of individual leaning (persistence) and neighbourhood leaning (persistence) for the nodes of the network. Also in this case the colour represents the density of users. Note that also marginal distribution are shown.
In more detail, panels (a) and (c) show the presence of echo chambers in the political and the fact-checking dimensions. Indeed, we note the presence of distinct areas in which users aggregate with others similar to them. Such areas are located on the diagonal of the plots indicating a positive correlation between individual leaning (persistence) and neighbourhood leaning (persistence). This homogeneous mixing could be, at least partially, driven by the homophilic tendency of users to interact with people that consume similar content both in terms of political bias and reliability.

To understand if echo chambers represent a peculiar feature of the empirical network that we are taking into account, we randomised the links of the initial bipartite network through the Maslov-Sneppen (MS) algorithm.
Such an algorithm was employed to obtain, after $2 x 10^8$ rewiring steps, a new randomised co-commenting network that has the same degree distribution of the original one. The MS algorithm is based on the simple principle of rewiring network links by switching their endpoints and it works as follows: i) sample a couple of links uniformly at random (e.g. A--B and C--D) ii) switch their endpoints (thus obtaining the new links A--D and B--C) iii) if at least one of the new links already exists abort the iteration step and select a new pair of links iv) repeat the procedure a number of times proportional to the number of links of the network.
The results are shown in panels (b) and (d) of Figure~\ref{fig:echo_chambers}, in which we may note that the echo chamber effect disappears. Thus, the presence of echo chambers can be considered a non-random topological feature of the empirical network. After the randomisation, users regardless their value of $p_i$ and $c_i$ are in contact with users with different values of those indexes thus resulting in an average distribution with little variation on the y axis, as shown by marginal plots.


To give a quantitative description of our results, we compute the joint entropy of the distributions shown in Figure~\ref{fig:echo_chambers}. The joint entropy of two discrete random variables is a measure of their degree of uncertainty. Consider two random variables $X$ and $Y$, the joint entropy is defined as:

\begin{equation}\label{je}
    H(X,Y) = - \sum_{x \in \mathcal{X}} \sum_{y \in \mathcal{Y}} P(x,y) \log_2 (P(x,y))
\end{equation}

where $X$ is the range of $X$, $Y$ is the range of $Y$ and $P(x, y)$ is the joint probability of those values.
If $P(x, y) = 0$ we assume that also $P(x, y) log_2(P(x, y)) = 0$.
The interpretation of Equation~\ref{je} relies on the concept of information content: entropy measures the average amount of information carried by the outcome of a trial to predict future outcomes and how uncertain the outcome is. The distribution with the highest entropy is the uniform distribution since there is no way to predict the future outcomes and it assumes the value $log_2(n)$, where $n$ is the number of possible couples $(x, y)$. On the other hand, the distribution with the lowest entropy value $(H(X, Y) = 0$ is $P (x, y) = \delta(x_0,y_0)$ (where $\delta(x_0, y_0)$ is the Dirac Delta function), since it is possible to predict exactly what the next outcome is.
We computed joint entropy for the joint distributions shown in Figure~\ref{fig:echo_chambers} and compared resulting entropy values with their random counterparts. To compute the joint probability we employed a quantization of the space using a grid with steps of 0.01 (used to sampling the frequencies of the distributions). We created a matrix of frequencies for each distribution. Since $p_i \in [-1, 1]$ and $c_i \in [0, 1]$ the matrices corresponding to the political distribution had size 200x200, while the matrices related to the persistence distribution had size 100x100. The values were then normalised by their maximum value $log_2(n)$. We obtain 0.3863 for the Political Bias and 0.2914 for its random counterpart. Instead, we obtain 0.4297 for the Persistence and 0.3338 for its random counterpart. 

The computed values show that echo chambers, by clustering the opinion in several distinct points of the space, have higher entropy values with respect to their random counterparts. This can be explained by noticing that, after the randomisation process, users interact with users in a wide spectrum of values and therefore the (average) Leaning/Persistence is centred in (approximately) one point, resulting in reduced entropy.

\section{Conclusion}

In this paper, we studied information diffusion during the COVID-19 pandemic by analysing the interaction of users with YouTube channels related to news outlets. 
Our findings show that, during the COVID-19 pandemic, the discussion was structured in echo chambers considering both the political and fact-checking dimensions. Furthermore, a substantial difference between the echo chambers behaviour and the random behaviour was highlighted in a quantitative manner.
A possible limitation of our study is the usage of a single source, namely Media Bias Fact Check whose rating methodology despite being public could be affected by subjective judgements, to label YouTube channels. This limitation could be overcome by considering further labelling available online or alternative methodologies to infer the political bias. 
%
%
Future work will involve both a deeper investigation of the network structure to describe the role of users in the process of commenting and a syntactic/semantic analysis of the text of such comments.   

\bibliography{aaai22.bib}

\begin{thebibliography}{11}
\providecommand{\natexlab}[1]{#1}
\providecommand{\url}[1]{\texttt{#1}}
\providecommand{\urlprefix}{URL }
\expandafter\ifx\csname urlstyle\endcsname\relax
  \providecommand{\doi}[1]{doi:\discretionary{}{}{}#1}\else
  \providecommand{\doi}{doi:\discretionary{}{}{}\begingroup
  \urlstyle{rm}\Url}\fi

\bibitem[{Bakshy, Messing, and Adamic(2015)}]{bakshy2015exposure}
Bakshy, E.; Messing, S.; and Adamic, L.~A. 2015.
\newblock Exposure to ideologically diverse news and opinion on Facebook.
\newblock \emph{Science} 348(6239): 1130--1132.

\bibitem[{bias/fact check(2014)}]{mbfc2014}
bias/fact check, M. 2014.
\newblock Media bias/fact check - search and learn the bias of news media.
\newblock \urlprefix\url{https://mediabiasfactcheck.com/}.

\bibitem[{Briand et~al.(2021)Briand, Cinelli, Nguyen, Lewis, Prybylski,
  Valensise, Colizza, Tozzi, Perra, Baronchelli et~al.}]{briand2021infodemics}
Briand, S.~C.; Cinelli, M.; Nguyen, T.; Lewis, R.; Prybylski, D.; Valensise,
  C.~M.; Colizza, V.; Tozzi, A.~E.; Perra, N.; Baronchelli, A.; et~al. 2021.
\newblock Infodemics: A new challenge for public health.
\newblock \emph{Cell} 184(25): 6010--6014.

\bibitem[{Cinelli et~al.(2021{\natexlab{a}})Cinelli, Morales, Galeazzi,
  Quattrociocchi, and Starnini}]{cinelli2021echo}
Cinelli, M.; Morales, G. D.~F.; Galeazzi, A.; Quattrociocchi, W.; and Starnini,
  M. 2021{\natexlab{a}}.
\newblock The echo chamber effect on social media.
\newblock \emph{Proceedings of the National Academy of Sciences} 118(9).

\bibitem[{Cinelli et~al.(2021{\natexlab{b}})Cinelli, Pelicon, Mozeti{\v{c}},
  Quattrociocchi, Novak, and Zollo}]{cinelli2021dynamics}
Cinelli, M.; Pelicon, A.; Mozeti{\v{c}}, I.; Quattrociocchi, W.; Novak, P.~K.;
  and Zollo, F. 2021{\natexlab{b}}.
\newblock Dynamics of online hate and misinformation.
\newblock \emph{Scientific reports} 11(1): 1--12.

\bibitem[{Feezell, Wagner, and Conroy(2021)}]{feezell2021exploring}
Feezell, J.~T.; Wagner, J.~K.; and Conroy, M. 2021.
\newblock Exploring the effects of algorithm-driven news sources on political
  behavior and polarization.
\newblock \emph{Computers in human behavior} 116: 106626.

\bibitem[{Garimella et~al.(2018)Garimella, Morales, Gionis, and
  Mathioudakis}]{garimella2018quantifying}
Garimella, K.; Morales, G. D.~F.; Gionis, A.; and Mathioudakis, M. 2018.
\newblock Quantifying controversy on social media.
\newblock \emph{ACM Transactions on Social Computing} 1(1): 1--27.

\bibitem[{Hosseinmardi et~al.(2021)Hosseinmardi, Ghasemian, Clauset, Mobius,
  Rothschild, and Watts}]{hosseinmardi2021examining}
Hosseinmardi, H.; Ghasemian, A.; Clauset, A.; Mobius, M.; Rothschild, D.~M.;
  and Watts, D.~J. 2021.
\newblock Examining the consumption of radical content on YouTube.
\newblock \emph{Proceedings of the National Academy of Sciences} 118(32).

\bibitem[{Ribeiro et~al.(2020)Ribeiro, Ottoni, West, Almeida, and
  Meira~Jr}]{ribeiro2020auditing}
Ribeiro, M.~H.; Ottoni, R.; West, R.; Almeida, V.~A.; and Meira~Jr, W. 2020.
\newblock Auditing radicalization pathways on YouTube.
\newblock In \emph{Proceedings of the 2020 conference on fairness,
  accountability, and transparency}, 131--141.

\bibitem[{Sharot and Sunstein(2020)}]{sharot2020people}
Sharot, T.; and Sunstein, C.~R. 2020.
\newblock How people decide what they want to know.
\newblock \emph{Nature Human Behaviour} 4(1): 14--19.

\bibitem[{Sunstein(2004)}]{sunstein2004democracy}
Sunstein, C.~R. 2004.
\newblock Democracy and filtering.
\newblock \emph{Communications of the ACM} 47(12): 57--59.

\end{thebibliography}

\end{document}